\documentclass[12pt,preprint]{aastex}

\usepackage{graphicx}

\slugcomment{AJ Revised 2009-February-19}

\shorttitle{P/2008 R1}
\shortauthors{Jewitt, Yang and Haghighipour}

\begin{document}


\title{Main-Belt Comet P/2008 R1 (Garradd)
 \footnote{
       Some data were obtained at the W. M.
       Keck Observatory,  a scientific partnership
       run by the California Institute of Technology, the University of
       California, and the National Aeronautics and Space Administration.
       The Observatory was made possible by the generous financial support
       of the W. M. Keck Foundation.  Some data were collected at the 
       Subaru Telescope, which is operated by the National Astronomical 
       Observatory of Japan.
}   
}

\author{David Jewitt, Bin Yang and Nader Haghighipour}
\affil{Institute for Astronomy, University of Hawaii, \\
2680 Woodlawn Drive, Honolulu, HI 96822}

\email{jewitt@ifa.hawaii.edu, yangbin@ifa.hawaii.edu, nader@ifa.hawaii.edu}



\begin{abstract}
We present a study of the newly-discovered main-belt comet P/2008 R1 (Garradd), an object with the dynamical characteristics of an asteroid and the physical characteristics of a comet.  Photometry sets a limit to the effective radius of the nucleus at $r_e <$ 0.7 km (red geometric albedo 0.05 assumed).  The coma shows a secular fading in our data caused by the escape of dust particles from the near-nucleus environment.  The optical reflection spectrum is a nearly neutral continuum devoid of gaseous emission lines, from which we derive a limit to the cyanide (CN) radical production rate of $Q_{CN} <$ 1.4$\times$10$^{23}$ s$^{-1}$ and infer a mass loss rate $<$ 1.5 kg s$^{-1}$ at the time of our observations.  Unlike the first-reported main-belt comets, P/2008 R1 is not dynamically stable.  The nearby 8:3 mean-motion resonance with Jupiter induces dynamical instability on timescales 20 to 30 Myr.   Hence, we conclude that P/2008 R1 has recently arrived from a more stable source elsewhere.  The high Tisserand parameter of the orbit (in fact, with $T_J$ = 3.216 it is the highest of any comet) points to a source in the asteroid belt itself, instead of in the Kuiper belt (putative source of the Jupiter family comets).   We infer that P/2008 R1 is an icy body from the outer asteroid belt in which sublimation has been triggered by rising temperatures resulting from a decreasing perihelion distance.
\end{abstract}

\keywords{minor planets, asteroids; comets: general; Kuiper Belt; solar system: formation}

\section{Introduction}

The Tisserand parameter with respect to Jupiter provides a simple way to discriminate dynamically between asteroids and comets (Kresak 1982, Kosai 1992).  It is defined by

\begin{equation}
T_J = \frac{a_J}{a} + 2\left[(1-e^2)\frac{a}{a_J}\right]^{1/2}\cos(i)
\label{tisserand}
\end{equation}

\noindent where $a$, $e$ and $i$ are the semimajor axis, eccentricity and inclination of the orbit and $a_J$ is the semimajor axis
of Jupiter.  Jupiter has $T_J$ = 3.0, the main-belt asteroids
have $T_J > $ 3 while comets have $T_J <$ 3.  The $T_J$ parameter is a constant of the motion in the circular, restricted three-body (Sun-Jupiter-object) problem (Carusi et al.\ 1995).   However, the assumptions of the circular, restricted three-body problem are not exactly met: Jupiter's orbit is inclined, is not circular and objects other than the Sun and Jupiter can exert significant dynamical influence on the motions of comets and asteroids.  Therefore, while Equation (\ref{tisserand}) is broadly useful, it can be ambiguous as a dynamical discriminant particularly for values of the parameter very close to $T_J$ = 3.0.   

Observationally, comets and asteroids are distinguished by the presence or absence of a coma (gravitationally unbound atmosphere) or tail.  A small body which shows a coma at any time is, by the observational definition, a comet (Jewitt 2004).  This observational distinction is also imperfect, because the observational ability to detect a coma is a function of the instrumental sensitivity and resolution.  Nevertheless, it is noteworthy that most objects classified dynamically as comets  by Equation (\ref{tisserand}) show distinctive cometary morphologies.  The opposite is also true: most objects which are dynamically asteroidal are also judged as asteroids by their point-like appearances.  This agreement between the dynamical and observational definitions reflects a sharp compositional difference between refractory objects formed at high temperatures in the inner regions of the protoplanetary disk and ice-rich bodies formed at larger distances beyond the snow-line.   

However, some objects which are dynamically comets ($T_J <$ 3) are perpetually devoid of comae and tails, the principal indicators of cometary activity.  Of these, some appear to be former main-belt asteroids that have escaped from their original orbits and are being multiply-scattered by Jupiter and the terrestrial planets into eccentric, comet-like orbits. Others are inferred to be inactive comets in which the near-surface supply of volatiles has been depleted by past sublimation losses (Hartmann et al 1987, Bottke et al.\ 2002).  In support of the latter inference is the observation that $T_J$ and albedo are correlated, such that inactive objects with $T_J <$ 3 are statistically more likely to have low, comet-like albedos than objects with $T_J >$ 3 (Fernandez et al.\ 2005).  Presumably, this is because a substantial fraction of the  $T_J <$ 3 ``asteroids''  are deactivated comets which retain their distinctive, low albedos. The ratio of original comets to original asteroids in the $T_J <$ 3 population is uncertain but may be near 50\% (Fernandez et al.\ 2005).

Recent work has revealed a second  and completely different class in which the dynamical and morphological classifications disagree (Hsieh and Jewitt 2006).  The ``main-belt comets'' (hereafter MBCs) have asteroid-like semimajor axes and Tisserand parameters, but they also display comet-like mass loss, manifested optically as resolved comae and tails.  
The prototype MBC is 133P/Elst-Pizarro ($T_J$ = 3.184, Hsieh et al.\ 2004) and two other examples, P/2005 U1 (Read) ($T_J$ = 3.153) and 176P/Linear ($T_J$ = 3.166), are well established (Hsieh and Jewitt  2006).  As with the ``normal'' comets from the Oort cloud and Kuiper belt, mass loss from the MBCs is compatible with being driven by the sublimation of near-surface ice and it is appropriate to think of the MBCs as ``icy asteroids''.   

It is very unlikely that the MBCs have been recently captured into the main-belt from sources in the Oort cloud or Kuiper belt (Fernandez et al.\ 2002, Levison et al.\ 2006).  The dynamical reason is that the aphelia lie so far from Jupiter that the MBCs are decoupled from the very object needed to effect their capture.   Some objects might have been captured from the Kuiper belt into the main-belt during a hypothesized, dynamically-chaotic phase about 0.8 Gyr after Solar system formation (Gomes et al.\ 2005).   However, the most likely explanation is that the MBCs formed in-situ at a time when the snow-line was closer to the Sun than $\sim$3 AU. Models suggest that this condition prevailed over a wide range of accretion parameters in the first few Myr of the Solar system (Garaud and Lin 2007).

Here, we present observations of the newly-discovered MBC P/2008 R1 (Garradd) starting only 2 days after its announcement on UT 2008 Sep 24 (Garradd et al.\ 2008).  Equation (\ref{tisserand}), with $a$ = 2.726 AU, $e$ = 0.342 and $i$ = 15.9$^{\circ}$, gives $T_J$ = 3.216, clearly above the nominal dividing line separating the comets and asteroids and higher even than the $T_J$ parameters of the other MBCs.   Based on these results, we initiated a program of optical observations of P/2008 R1 that form the basis of the current work.

\section{Observations} 
Observations were taken using three telescopes located on Mauna Kea, Hawaii.  The University of Hawaii 2.2-m telescope was used with a Tektronix 2048$\times$2048 pixel charge-coupled device (CCD) camera, an image scale of 0.$\!\!^{\prime\prime}$219 pixel$^{-1}$ and $BVRI$ filters approximating the Kron-Cousins set. At the Keck 10-m telescope we used the two-channel LRIS imaging camera (Oke et al.\ 1995) with scale 0.$\!\!^{\prime\prime}$211 pixel$^{-1}$ on the red side ($VRI$ filters) and 0.$\!\!^{\prime\prime}$135 pixel$^{-1}$ on the blue side.  Spectra were also secured at Keck by using LRIS in its dispersive mode.  The Subaru 8-m telescope was used with the FOCAS 4096$\times$4096 pixel CCD imager binned 2$\times$2 to give an image scale 0.$\!\!^{\prime\prime}$208 pixel$^{-1}$ through $BVRI$ filters (Kashikawa et al.\ 2002).  The observations were taken with the telescopes tracked at non-sidereal rates to follow the motion of P/2008 R1, using a range of integration times from 60 s to 300 s with multiple images in each filter.   All images were photometrically calibrated using observations of nearby standard stars having Sun-like optical colors from the list by Landolt (1992).  Images from UT 2008 Sep.\ 26 were taken through cloud and calibrated using subsequent observations of field stars taken on a photometric night.  The other nights were photometric to within $\sim$0.05 mag., or better.  The low declination of P/2008 R1, especially before mid-October, severely limited the nightly observing window and forced observations to be taken at larger than normal airmasses (typically 2.0 - 2.2).

A journal of observations is given in Table (\ref{geometry}), while an image showing the cometary appearance on UT 2008 Sep.\ 30 is given in Figure (1).   A tail extends primarily along the projected Sun-comet radius vector and has the diffuse appearance expected of a particle tail swept back from the nucleus by radiation pressure.  The comet retained the appearance shown in Figure (1) throughout the observations reported here.

\section{Analysis}
\subsection{Photometry} 

We took photometry using a circular projected aperture of 2.$\!\!^{\prime\prime}$2 angular radius, and measured the sky brightness from a surrounding annulus having inner and outer radii of 4.$\!\!^{\prime\prime}$4 and 8.$\!\!^{\prime\prime}$8, respectively.  We found by experimentation that substantially smaller sky annuli overlap the coma unacceptably while substantially larger annuli give less reliable estimates of the sky brightness in the vicinity of the comet.  The comet tail crosses the sky annulus.  We checked that, since we employ the median pixel value within the annulus as our best estimate of the sky, the resulting sky brightness is unaffected by the comet tail. Images in which the comet was overlapped by field stars and galaxies were rejected from further analysis.  The apparent magnitudes (averages of several measurements per filter per night), are listed in column 2 of Table \ref{phot}.


The apparent brightness of P/2008 R1 (Table \ref{phot}) fades by about 1.5 mag. from the first observation to the last.  To account for the effects of viewing geometry in the photometry, we corrected the apparent magnitudes of the comet to unit heliocentric and geocentric distances and to zero degrees phase angle.  We assumed an $R^{-2} \Delta^{-2}$ inverse-square law brightness variation, as is strictly applicable to photometry of a point source.  However, P/2008 R1 is an extended object and the volume of the coma sampled by fixed-aperture photometry (as in Table \ref{phot}) grows with increasing geocentric distance, tending to make measurements at large $\Delta$ too bright relative to ones at small $\Delta$, all else being equal (Jewitt 1991).  The magnitude of this ``aperture effect''  depends upon the radial variation of the surface brightness (Jewitt 1991).    For example, we estimate an additional $\sim$0.3 mag. fading in $H_R$ between the first and last observations (c.f.\ Table \ref{phot}), tending to steepen the line in Figure (2).  The important point is that this aperture effect always increases the amount of fading so that the latter cannot be an artifact of the presence of coma.  We next used the ``HG'' formalism (Bowell et al.\ 1989)  to correct for the phase-angle 
dependence of the scattered flux density, with the dimensionless 
scattering parameter  $G$ = 0.15 as appropriate for C-type objects.   
The correction to 0$^{\circ}$ phase angle with this $G$ is $\Delta m_R$ = 1.27 mag.\ for $G$ = 0.15.  If we had chosen instead to assume $G$ = 0.25 (more appropriate for a high albedo, S-type surface) we would obtain a phase correction $\Delta m_R$ = 1.14 mag.  The difference is not important relative to the size of the variations in brightness recorded in Table 2.  Of course, P/2008 R1 is an extended object and there is no guarantee that its phase function is well-represented by the ``HG'' model, regardless of the $G$-value selected.   Empirically, the back-scattering phase functions of active comets are approximately linear with gradients 0.02 mag.\ deg$^{-1}$ to 0.04 mag.\ deg$^{-1}$ (Millis et al. 1982, Meech and Jewitt 1987).  For P/2008 R1 (Table \ref{geometry}) these gradients correspond to phase corrections $\Delta m_R$ = 
0.6 to 1.2 mag.\ giving a measure of the uncertainty implicit in the phase correction.  For definiteness, in this work we assumed  $G$ = 0.15, consistent with the low albedos derived from optical/thermal infrared measurements of MBCs 133P (red geometric albedo $p_R$ = 0.05$\pm$0.02) and 176P ($p_R$ = 0.07$\pm$0.02; Hsieh et al.\ 2009) and we note that the phase correction could be smaller by $\sim$0.6 mag., corresponding to a factor of $\sim$1.7.  Resulting values of $H_R$ are listed in column 3 of Table \ref{phot}.  

The coma of P/2008 R1 limits our ability to study the nucleus directly.  In principle, the surface brightness of the coma can
be extrapolated inwards to the nucleus and then subtracted from the data to estimate the nucleus contribution.  In practice, this exercise is fraught with uncertainty.  In particular, reliable results cannot be obtained when the combined cross-sections of the dust grains within the projected seeing disk are large compared to the geometrical cross-section of the nucleus.  An examination of the surface brightness profile of P/2008 R1 shows no clear evidence for any central excess that can be unambiguously associated with the presence of an underlying nucleus.   For this reason, we have not attempted to constrain the nucleus size using the surface brightness profile.  For the same reason, we can place no meaningful constraint on the shape of the nucleus or its rotational period based on our data.

Instead, we use the faintest absolute magnitude from Table \ref{phot}, namely $H_R$ = 17.77, to set a limit to the absolute nuclear magnitude.  The inverse-square law 

\begin{equation}
p_R r_e^2 = 2.25\times10^{16} 10^{0.4[m_{\odot} - H_R]}
\end{equation}

\noindent where $m_{\odot}$ = -27.1 is the $R$-band magnitude of the Sun, then gives an upper limit to the effective radius of the nucleus, $r_e$ [km], as 

\begin{equation}
r_e < 0.7 (0.05/p_R)^{1/2}. 
\end{equation}

\noindent A $\sim$0.6 mag.\ fainter absolute magnitude (caused by a significantly flatter phase correction, as above) would result in a derived radius about 30\% smaller, or $r_e \sim$ 0.5 km.

The absolute magnitude shows a secular decline of about 0.5 mag.\ (i.e.\ a fading of about 40\%) from 
the first observation to the last (Figure (2) and Table \ref{phot}).  The fading results from a net loss of near-nucleus dust particles projected within the photometry aperture, presumably due to radiation pressure sweeping (the 2.$\!\!^{\prime\prime}$2 aperture radius corresponds to about 1800 km at 1.1 AU).   Unfortunately, the Earth never rises far above the plane of the orbit of P/2008 R1 (on UT 2008 Sep 30, this angle was 9.8$^{\circ}$) leading to strong projection effects that limit the useful application of Finson-Probstein type dust models.  

Instead, we obtain a crude estimate of the particle coma mass loss rate from the photometry, as follows.  The largest and smallest effective radii in Table \ref{phot} are $r_e(max)$ = 1.03 km (UT 2008 Sep 26) and $r_e(min)$ = 0.71 km (UT 2008 Oct 22), respectively.  The loss in dust cross-section between these dates amounts to $\Delta C$ = $\pi (r_e(max)^2 - r_e(min)^2)$ = 1.75 km$^2$, assuming that the dust particles in the coma have the same albedo and scattering function as the nucleus).  For spherical particles of density $\rho$ [kg m$^{-3}$] and average radius $a$ [m], the mass and cross-section are related by $\Delta M \sim \rho a \Delta C$ [kg].  With $a$ = 10 $\mu$m (Hsieh et al. 2004) and $\rho$ = 10$^3$ kg m$^{-3}$, we find that the coma lost $\Delta M$ = 1.8$\times$10$^4$ kg between Sep 26 and Oct 22, corresponding to an average loss rate of $\sim$10$^{-2}$ kg s$^{-1}$.   For comparison, mass loss rates estimated from detailed modeling of the MBC prototype 133P are (1 to 10)$\times$10$^{-3}$ kg s$^{-1}$ (Hsieh et al. 2004) while most active comets are 3 to 6 orders of magnitude more productive.  Strictly, our estimate gives only the dust mass loss rate from the coma (i.e.\ the difference between the rate of supply from the nucleus and the rate of loss by dissipation into space) so it is difficult to attach an estimate of uncertainty to it.   An independent constraint on the gas production rate, given in the next section, depends upon fewer assumptions and is correspondingly more robust.

The optical colors of P/2008 R1 are given in Table \ref{phot}, measured within a 2.$\!\!^{\prime\prime}$2 radius aperture using an observing cadence designed to mitigate against the possible effects of rotational brightness variations.  Within the measurement uncertainties, the colors from different nights are in broad agreement except that the $B-V$ measurement from UT 2008 Sep 30 is significantly redder than $B-V$ on later dates.  We can find no reason to reject this measurement.  Therefore, we have retained all the data in the Table in the computation of the weighted mean colors within each band, listed in Table \ref{phot}, together with the colors of the Sun from Hardorp (1982) and Hartmann et al.\ (1990). The mean and Solar colors are the same within the errors of measurement and consistent with the nearly neutral reflection spectra of C-type asteroids. The mean colors are inconsistent with the redder P- and D-types that are common in the Trojan asteroids (Fornasier et al.\ 2007, Roig et al.\ 2008) and there is no evidence for the ultrared matter observed to coat the surfaces of many Kuiper belt objects and Centaurs (Jewitt 2002).   The physical interpretation of the colors is limited, however, by the knowledge that they refer primarily to  scattering from near-nucleus dust and not from the underlying nucleus.  The scattering properties of coma dust grains can, in principle, be quite different from those of the solid nucleus surface, especially if the particles are small compared to the wavelength.  We searched for spatial gradients in the colors that might provide evidence for a nucleus different in color from the surrounding coma. In the Keck data from UT 2008 Sep 30 and the UH 2.2-m data from UT 2008 Oct 03 no spatial variations in color were found.  The optical continuum reflectivity gradients of nine active comets vary from 5$\pm$2 to 18$\pm$2 \%/1000\AA~(Jewitt and Meech 1986), corresponding to $V-R$ = 
0.41 to 0.54. The $V-R$ color of P/2008 R1 is near the bluest of the measured cometary coma colors but is not otherwise remarkable. 


\subsection{Spectrum}
A spectrum was recorded at the Keck telescope on UT 2008 Sep 30 in order to search for evidence of emission lines from gas.  To obtain the spectrum, P/2008 R1 was first identified in the Keck guider camera from its distinctive non-sidereal motion and then placed within a 1.$\!\!^{\prime\prime}$0 wide slit.  Spectral dispersion was obtained using the 400/3400 grism with the 460 dichroic, giving a dispersion of 1.07 \AA~pixel$^{-1}$ and a resolution of approximately 5 \AA.  An atmospheric dispersion compensator was used and the orientation of the slit was further held fixed perpendicular to the horizon in order to obviate effects from differential atmospheric refraction.  A total of 1710 s of data were obtained from four integrations on the comet.  Three Solar analog (G2V-type) reference stars (HD202561, HD206416  and HD211132) were observed in order to cancel features from the photospheric spectrum of the Sun.   We selected these stars on the basis of their (nominal) G2V classifications but also on the basis of their proximity in the sky to the target MBC.  Owing to its large negative declination (-41$^{\circ}$) at the time of observation, the comet spectra were necessarily obtained at airmasses 2.05 - 2.08.  The G-type stars were observed immediately after P/2008 R1 at airmasses in the range 2.03 - 2.20.   Therefore, effects of differential extinction between the comet and the stars should be negligible.   Wavelength calibration and quartz lamp flat-fields were taken for calibration purposes.

The $\sim$3700\AA~to $\sim$4300\AA~portion of the spectral image, prior to sky-subtraction, is shown in Figure (3).    The cometary continuum is evident as a horizontal band crossed by emission lines from the night sky and by absorption bands in the spectrum of the Sun, visible in reflection from dust in the Zodiacal cloud.   The latter include the prominent Ca H and K lines at 3933\AA~and 3966\AA, marked near the middle of Figure (3).   Cometary emission bands, if they were present, would appear in Figure (3) as vertical bands brightest at the location of the continuum and fading with distance above and below it.   The brightest expected cometary species in this wavelength range is the CN $\Delta$v = 0 band, which occupies a range of wavelengths from roughly 3830\AA~to the bandhead at 3883\AA.  Neither CN $\Delta$v = 0 nor any other cometary bands are evident. 

Figure (4) shows the spectrum extracted from the data using a section of the slit centered on the continuum and 3.$\!\!^{\prime\prime}$8 in length, with the sky emission subtracted from flanking boxes extending 5$''$ to 8$''$ from the comet.  The extracted spectrum has been divided by the spectrum of G2V star HD202561 and the result has been further divided by a linear function in order to remove a slight residual slope (of unknown origin, but possibly resulting from an imperfect analog) in the computed spectral ratio.  Although the Ca H and K lines (locations are marked in Figure 4) are neatly cancelled by the division by the G-type star, residuals in the ratio spectrum, especially near 3820\AA, show that HD202561 is not a perfect analog of the Sun.  The other two G-stars also gave imperfect line cancellations, with resulting ratio spectra similar to the one shown in the Figure.   The CN line is not apparent in the extracted spectrum of P/2008 R1.

We estimate a limit to the flux density in the CN band (Figure 4) as follows.  The $B$ filter (central wavelength 4500\AA) magnitude of P/2008 R1 is $m_B$ = 21.02$\pm$0.03 (Table \ref{phot}), which corresponds to a flux density at band center of $f_B$ = 2.5$\times$10$^{-17}$ erg cm$^{-2}$ s$^{-1}$ \AA$^{-1}$ (Drilling and Landolt 2000).   The reflectivity spectrum of P/2008 R1 is essentially independent of wavelength (as shown by the nearly neutral broadband colors in Table 2).  Therefore, the continuum flux density of P/2008 R1 at the wavelength of CN $\Delta$v = 0 is just
 $f_{CN}^C = f_B [f_{3880}/f_B]_{\odot}$ in which [$f_{3880}/f_B]_{\odot}$ = 0.52 is the ratio of the flux densities in the Solar spectrum at the wavelengths of CN  and the $B$-filter center (Arvesen et al.\ 1969).  This gives  $f_{CN}^C$ = 1.3$\times$10$^{-17}$ erg cm$^{-2}$ s$^{-1}$ \AA$^{-1}$.  

To quantify a limit to the possible strength of the CN band we examined the standard error on the mean of the ratio spectrum in Figure (4) over 50\AA~wavelength bins on either side of the expected CN emission band.  In the wavelength bands from 3780\AA~to 3830\AA , 3900\AA~to 3950\AA~and 3950\AA~to 4100\AA~we measured the standard error on the mean of the normalized reflectivity (Figure 4) as $\sigma$ = 0.06 , 0.04 and 0.03, respectively.  To be conservative, we select the largest ($\sigma$ = 0.06) as the best measure of the uncertainty in the data and consider the practical limit to the CN band flux as 5$\sigma$ = 0.30.  This limit is plotted as a horizontal line in Figure (4).  Expressed as a flux in the $\Delta \lambda$ = 50\AA~wide band occupied by CN, this limit becomes  $f_{CN}$ = 0.3 $\Delta \lambda f_{CN}^C$ = 2.0$\times$10$^{-16}$ erg cm$^{-2}$ s$^{-1}$ and we employ this value to calculate the CN production rate limit.   

The CN band is produced by resonance fluorescence of sunlight, in which cometary molecules scatter photons randomly in direction but without change in energy.  When the coma is optically thin, the scattered flux, $f_{CN}$ [erg cm$^{-2}$ s$^{-1}$], received at distance $\Delta$ [cm] is simply proportional to the number of CN molecules, $N$, in the section of the slit from which the spectrum was extracted, or

\begin{equation}
f_{CN} = \frac{g(R) N}{4 \pi \Delta^2}.
\label{fluorescence}
\end{equation}

\noindent The constant of proportionality in Equation \ref{fluorescence} is called the resonance fluorescence efficiency, $g(R)$ [erg s$^{-1}$ molecule$^{-1}$].  It is conventionally evaluated at heliocentric distance $R$ = 1 AU and scaled to other distances using $g(R)$ = $g(1)$/$R^2$.  At a given $R$, the fluorescence efficiency of CN can vary by a factor of $\sim$1.5, owing to the Doppler shift between cometary lines and the Solar spectrum (the so-called Swings effect).   On UT 2008 Sep 30, P/2008 R1 had radial velocity $dR$/$dt$ = 3.4 km s$^{-1}$, for which the fluorescence efficiency is $g(1)$  = 2.8$\times$10$^{-13}$ [erg s$^{-1}$ molecule$^{-1}$] (work by M. Mumma quoted in A'Hearn 1982). 

Substituting $f_{CN}$ = 2$\times$10$^{-16}$ erg cm$^{-2}$ s$^{-1}$, and with $R$ and $\Delta$ drawn from Table \ref{geometry} for UT 2008 Sep 30, we obtain $N$ = 8.4$\times$10$^{24}$ CN molecules in the 1.0$\times$3.$\!\!^{\prime\prime}$8 projected slit.  To estimate the production rate from emission line data we used the Haser model, in which the parent of CN decays with a length scale (at $R$ = 1 AU) $\ell_P$ = 1.3$\times$10$^4$ km and the CN daughter itself is photo-destroyed on the length scale $\ell_{d}$ = 2.1$\times$10$^5$ km (A'Hearn et al.\ 1995).  We integrated the Haser number density model over a rectangular area 820 km $\times$ 3940 km, which is the size of the spectrograph slit projected to the distance of P/2008 R1.  We assumed that the outflow speed of the sublimated gas is $v$ = 0.75 km s$^{-1}$, consistent with measurements in comet C/Hale-Bopp at about this heliocentric distance (Biver et al.\ 2002). The resulting limit to the production rate of CN is $Q_{CN} <$ 1.4$\times$10$^{23}$ s$^{-1}$. 

In other comets, the average ratios of the production rates of CN to OH and of OH to H$_2$O are $Q_{OH}$/$Q_{CN} \sim$ 320 and $Q_{OH}$/$Q_{H_2O}$ = 0.9, respectively, giving $Q_{H_2O}$/$Q_{CN} \sim$ 360 (A'Hearn et al.\ 1995).  If this ratio applies to P/2008 R1, our CN limit implies $Q_{H_2O} <$ 5$\times$10$^{25}$ s$^{-1}$, corresponding to $<$ 1.5 kg s$^{-1}$.  This estimate, which is crude but probably good to an order of magnitude, lies at the low end of the range of CN production rates of other well-measured comets (A'Hearn et al.\ 1995) but is consistent with the dust mass loss rate earlier estimated from the photometry.   

The non-detection of gas in an active comet is unsurprising in the context of observations of other comets taken in the 2 to 3 AU region and beyond.  In fact, the visibility of spectral lines, relative to the dust continuum, declines systematically as the heliocentric distance increases. Many active Jupiter family comets (JFCs), for example, show prominent dust comae but lack measurable gas emission, just as in the case of P/2008 R1.  This is in part because the scattering cross-section of small dust particles is intrinsically large but also because the residence time of dust (leaving the nucleus at speeds 1 to 10 m s$^{-1}$) is 10$^2$ to 10$^3$ times larger than the residence time of gas (speeds $\sim$1 km s$^{-1}$), when measured within an aperture of fixed radius.  This difference in the residence times of gas and dust is potentially even more important in P/2008 R1, where the observed secular fading proves that a non-steady state prevails.  Our observations are consistent with a scenario in which most of the gas has already left the vicinity of the nucleus, leaving behind a dissipating coma consisting largely of slow-moving dust particles.

\subsection{Dynamics}

As is obvious in Figure (5), the orbit of P/2008 R1 is quite distinct from the orbits of the first three MBCs in having a smaller semimajor axis (2.7 AU vs.\ 3.2 AU), a larger eccentricity (0.35 vs.\ 0.2) and a much larger inclination (16$^{\circ}$ vs.\ $\sim$1$^{\circ}$).  We explored the stability of P/2008 R1 by integrating its orbit using the Bulirsch-Stoer integrator in the Mercury N-body integration package (Chambers 1999). We found that the orbit of P/2008 R1 becomes unstable in approximately 20 Myr, a time very short compared to the 4.5 Gyr age of the Solar system.  This dynamical instability is unsurprising because the semimajor axis of P/2008 R1 (2.727 AU) lies close to the 8:3 mean-motion resonance with Jupiter (at 2.706 AU, see Figure 6) and the orbit is also affected by the $\nu_6$ secular resonance, in a region that is known to be dynamically unstable (Yoshikawa 1989, Broz and Vkrouhlicky 2008).

To obtain a more comprehensive picture of the stability of P/2008 R1, we further integrated the motions of objects (clones) in the vicinity of this asteroid.  We computed future orbits over a grid of orbital elements in semimajor axis vs. eccentricity vs. inclination space (shown as a small box in Figure 5).  Results for a smaller number of clones are shown in Figure (6).  As shown here, the majority of clones became unstable with lifetimes between 20 Myr and 100 Myr implying that the region around P/2008 R1 is naturally unstable. In most of our simulations, objects were lost by excitation to high eccentricity due to interaction with the 8:3 mean-motion resonance. The integrations showed that the median lifetimes of a set of clones with inclinations and other elements similar to those of P/2008 R1 are 20 to 30 Myr.  This is shorter even than the $\sim$1 Gyr collisional lifetime of a kilometer sized asteroid in the main-belt.

The orbits of MBCs 133P and 176P are very similar to each other (see Table 3 and Figure 5) and to the orbits of the 2 Gyr-old Themis asteroid family (Zappala et al.\ 1990) and the $\sim$10 Myr old (and much smaller) Beagle family (Nesvorny et al.\ 2008).     The orbital resemblance between 133P and 176P is unsurprising, since 176P was discovered in a survey (Hsieh and Jewitt 2006) that specifically targeted objects having orbits like that of 133P.  The third MBC, P/2005 U1, was found by an unrelated survey that is not predisposed to discover 133P-like objects.  It falls suspiciously close to, but distinctly outside, the domains of the Themis and Beagle families. Furthermore, whereas 133P and 176P are dynamically stable on Gyr timescales, P/2005 U1 appears to be a dynamically short-lived object (Haghighipour 2009) as is P/2008 R1. It is possible that this MBC was formed in another location in the asteroid belt and reached its present orbit through interactions with several mean-motion resonances.  The small size of P/2008 R1 suggests that further exploration of the dynamics should probably take account of the Yarkovsky and YORP radiation forces (Bottke et al.\ 2006).  Indeed, depending on the exact size of the nucleus (for which we possess only an upper limit), radiation forces might play a defining role in driving the dynamical evolution.

\section{Discussion}

Water ice is thermodynamically unstable on the surface of P/2008 R1.  The isothermal blackbody temperature of a sphere placed at the perihelion distance ($q$ = 1.793 AU) is $T_{BB}$ = 208 K, falling to $T_{BB}$ = 145 K at aphelion ($Q$ = 3.658 AU).  If the nucleus rotates slowly, and/or if the spin axis intersects the Sun, temperatures at the subsolar point could potentially be $\sim$2$^{1/2}$ times higher, or about 294 K and 205 K at perihelion and aphelion, respectively.  These temperatures are high enough to guarantee that the sublimation of exposed, dirty water ice is important, but the magnitude of the sublimation rate is difficult to calculate with accuracy in the absence of more detailed knowledge about the body.  The sublimation rate depends exponentially on the temperature, and is therefore sensitive to the magnitude and direction of the spin vector as well as to the optical and thermophysical parameters of the surface.  A crude model, in which the sublimation occurs in thermal equilibrium and is averaged around the orbit (e.g.\ see Figure (6) of Jewitt 2004), gives a mean sublimation rate $dm$/$dt \sim$ 1.2$\times$10$^{-5}$ kg m$^{-2}$ s$^{-1}$.   It is clear that sublimation at this high rate cannot be sustained for long times.  Sublimation at rate $dm$/$dt$ corresponds to recession of the sublimation surface at rate $dr_e$/$dt$ = $\rho^{-1}$ $dm$/$dt \sim$ 1  mm day$^{-1}$, for bulk densities $\rho$ = 1000 kg m$^{-3}$.  In a single orbit, the recession of the sublimating surface would be $\Delta r_e \sim$ 1.6 m. The free sublimation lifetime is $t_s \sim r_e [dr_e/dt]^{-1}$.  With $r_e <$ 500 m, we find $t_e <$ 300 yr or about 70 orbits.  The inevitable conclusion is that ice on P/2008 R1 has only recently been exposed to the heat of the Sun.

The instability of surface ice and the short dynamical lifetime together indicate that P/2008 R1 was recently injected into its present orbit from a source region elsewhere.  Unfortunately, dynamical chaos prevents us from integrating backwards to determine the location of the source.  However, the large $T_J$ = 3.2 tends to associate P/2008 R1 with the asteroid belt, not with the Kuiper belt from which the typical JFCs are supplied.  The need to keep near-surface ice stable against sublimation favors source regions in the colder, outer asteroid belt.   The trigger for cometary activity in P/2008 R1 might then be plausibly associated with the thermal shock resulting from the higher Solar insolation caused by a decrease in the perihelion distance.  Near-surface volatiles previously protected from the Sun by a $\sim$meter thick layer of refractory debris (Schorghofer 2008) would sublimate at the elevated temperatures produced by the decreasing perihelion distance.  For example, if P/2008 R1 started with a nearly circular orbit at 3.2 AU, like that of 133P (Table 3), the insolation at perihelion would increase by a factor $\sim$2 and the radiation equilibrium temperature by a factor $\sim$2$^{1/4} \sim$1.2 (a temperature change of 30 K to 40 K).  The actual temperature change would be smaller, owing to buffering by increased sublimation rates.  Still, given the exponential dependence of the sublimation flux on the temperature, the inward drift of the perihelion could easily increase the gas pressure by several orders of magnitude and cause the destabilization of any mantle material overlying the ice.  Exactly this kind of coupling between decreasing perihelion distance and the regeneration of cometary activity is observed in the nuclei of short-period comets (Rickman et al.\ 1991) and it seems reasonable that it should also occur in an ice-rich asteroid displaced inwards.  

P/2008 R1 is unlikely to be unique, meaning that other asteroids in the middle and outer regions of the belt probably contain near-surface ice.  Indeed, buried ice may be ubiquitous in the outer belt, only becoming detectable when uncovered by small impacts (as suggested for the dynamically stable, low eccentricity MBCs like 133P: Hsieh and Jewitt 2006; Haghighipour 2009) or by thermal instability triggered by an inward-drifting perihelion, as in P/2008 R1.  Parallel evidence comes from independent spectroscopic work showing that the degree of hydration of silicate minerals in asteroids decreases as distance from the Sun increases, presumably because trapped ice in distant objects never melted (Jones et al.\ 1990).  It is difficult to infer the population of such icy asteroids, however, given the many biases present in the existing data, but the biases mostly act to hide ice and thus to support the conjecture that icy asteroids are common. For example, given its small size, it seems likely that the nucleus of P/2008 R1 would have completely escaped detection had it not been in an active state.  Similar bodies in which the ice lies buried only slightly deeper, or objects in which the spin vector is oriented to maintain a lower perihelion temperature, might never become sufficiently active to be noticed, further biasing the sample.   It is difficult to model these effects in the absence of reliable data about the structural, thermal and spin parameters of the asteroids.

It is tempting to try to link together the four known MBCs as products of a shattering collision, perhaps to be identified with the event responsible for the creation of the Beagle family, involving an icy parent body in the outer belt.   Some fragments from this collision (like 133P and 176P) would remain in stable orbits in the outer belt.  Others would be deflected by planetary perturbations and driven into short-lived, excited orbits (like P/2005 U1 and now P/2008 R1).   Conceivably, such an origin might be stretched to explain at least some of the Encke-type comets (which have $T_J$ very slightly larger than 3.0).  Published attempts to account for the orbit of comet 2P/Encke ($T_J$ = 3.03) assume that it is a JFC from an original Kuiper belt source that has somehow been decoupled from Jupiter.  Planetary perturbations and non-gravitational forces have been invoked for this decoupling and to drive the Tisserand parameter above 3.0 from starting JFC values $T_J <$ 3.  Both mechanisms are problematic. Non-gravitational force models necessarily assume sustained outgassing forces that seem extreme (and which would be vastly more effective on nuclei smaller than P/Encke's (Fernandez et al.\ 2002), leading to an over-abundance of small Encke-type objects that is not observed). Pure gravitational capture takes longer than the expected physical lifetime of P/Encke (Levison et al.\ 2006). However, we know of no evidence requiring that Encke-type comets originated in the Kuiper belt.  Perhaps some of the dynamical problems in capturing the Encke-type comets could be relieved if they started as original members of the outer main-belt and drifted inwards through gravitational interactions with the planets.

\section{Summary}

Main belt comet P/2008 R1 has a coma and tail like a comet but has a semimajor axis in the main asteroid belt and an asteroid-like Tisserand parameter with respect to Jupiter, $T_J$ = 3.216.   We find that:

\begin{enumerate}

\item Optical photometry sets an upper limit to the effective radius of the nucleus of P/2008 R1 equal to 0.7 km, assuming a red geometric albedo 0.05. 

\item Spectra show no trace of CN $\Delta V$ = 0 or other resonance fluorescence lines from gas.  The derived upper limits to gaseous production are $Q_{CN} <$ 1.4$\times$10$^{23}$ s$^{-1}$ which, with a canonical CN/water ratio measured in other comets, indicates a water production rate less than 5$\times$10$^{25}$ s$^{-1}$, or $<$ 1.5 kg s$^{-1}$, on UT 2008 Sept 30.

\item The photometry shows a secular fading over the UT 2008 Sept 26 to Nov 11 period, corresponding to the loss of dust from the near-nucleus coma at $\sim$10$^{-2}$ kg s$^{-1}$.

\item The optical colors of P/2008 R1 are Sun-like, with $B-V$ = 0.63$\pm$0.06, $V-R$ = 0.37$\pm$0.03 and $R-I$ = 0.39$\pm$0.03.  These  values are consistent with the colors of $C-$class asteroids that are known to dominate the outer belt.  However, the colors primarily reflect scattering from dust particles in the coma and may not represent the spectrum the underlying nucleus.

\item The dynamical lifetime is limited by planetary perturbations to only $\sim$20 to 30 Myr, meaning that P/2008 R1 must have recently arrived from elsewhere.  The short lifetime and large Tisserand parameter suggest that P/2008 R1 is an ice-carrying asteroid recently escaped from a longer-lived orbit elsewhere in the main belt.

\item The available data on the main-belt comets, although limited, are consistent with the hypothesis that a large fraction of outer belt ``asteroids'' contain buried ice.

\end{enumerate}

\acknowledgments
We thank Nuno Peixinho and Nick Moskovitz for offering observational assistance and
 telescope operators Dan Birchall (UH), Bob Potter (Subaru) and Terry Stickel (Keck) for
their help.  Henry Hsieh, Pedro Lacerda, Jing Li and the referee, Beatrice Mueller, provided 
helpful comments.  NH appreciates support 
from the NASA Astrobiology Institute under Cooperative 
Agreement NNA04CC08A, the office of the Chancellor of the University of Hawaii and a 
Theodore Dunham J. grant administered by Funds for Astrophysics 
Research, Inc.  We appreciate support from a NASA Planetary Astronomy grant to DJ.

\begin{deluxetable}{llllll}
\tabletypesize{\scriptsize}
\tablecaption{Observational Geometry 
\label{geometry}}
\tablewidth{0pt}
\tablehead{
\colhead{UT Date} & \colhead{$R$ [AU]\tablenotemark{a}} & \colhead{$\Delta$ [AU] \tablenotemark{b}}   & \colhead{$\alpha$ [deg]\tablenotemark{c}} & \colhead{Tel\tablenotemark{d}} & \colhead{Work\tablenotemark{e}}  }
\startdata
2008 Sep 26 & 1.853 & 1.092 & 26.5 & UH 2.2-m & R \\
2008 Sep 30 & 1.861 & 1.125 & 27.1 & Keck 10-m & BVRI, Sp \\
2008 Sep 30 & 1.861 & 1.125 & 27.1 & UH 2.2-m & BVR  \\
2008 Oct 01 & 1.863 & 1.133 & 27.3 & UH 2.2-m  & BVR\\
2008 Oct 02 & 1.865 & 1.142 & 27.4 & UH 2.2-m & BVRI \\
2008 Oct 03 & 1.867 & 1.150 & 27.6 & UH 2.2-m  & BVRI\\
2008 Oct 22 & 1.910 & 1.334 & 29.5 & Subaru 8-m & BVRI \\
2008 Nov 11 & 1.963 & 1.562 & 30.0 & UH 2.2-m & R \\

\enddata


\tablenotetext{a}{Heliocentric distance in AU at the mid-time of the observations.}
\tablenotetext{b}{Geocentric distance in AU at the mid-time of the observations.}
\tablenotetext{c}{Phase angle [degrees] at the mid-time of the observations.}
\tablenotetext{d}{Telescope}
\tablenotetext{e}{BVRI denotes broadband imaging, Sp denotes optical spectra}

\end{deluxetable}

\clearpage 

\begin{deluxetable}{lllllll}
\tabletypesize{\scriptsize}
\tablecaption{Optical Photometry 
\label{phot}}
\tablewidth{0pt}
\tablehead{
\colhead{UT Date} & \colhead{$m_R$\tablenotemark{a}} & \colhead{$H_R$\tablenotemark{b}}& \colhead{$r_e [km]$\tablenotemark{c}}   & \colhead{$B-V$} & \colhead{$V-R$} & \colhead{$R-I$}  }
\startdata

2008 Sep 26 & 19.69$\pm$0.05 & 16.96 & 1.03  & -- & --  & --\\
2008 Sep 30 & 19.88$\pm$0.03 & 17.06 & 0.99  & 0.81$\pm$0.05 & 0.33$\pm$0.05  & 0.35$\pm$0.05\\
2008 Sep 30 & 19.93$\pm$0.03 & 17.11 & 0.96 & -- & --  & --\\
2008 Oct 01 & 19.86$\pm$0.02 & 17.02  & 1.00 & - & --  & --\\
2008 Oct 02 & 19.90$\pm$0.03 & 17.04 & 0.99  & 0.60$\pm$0.05 & 0.42$\pm$0.05  & 0.41$\pm$0.05\\
2008 Oct 03 & 20.17$\pm$0.06 & 17.28 & 0.89 & 0.62$\pm$0.07 & 0.41$\pm$0.05  & 0.52$\pm$0.07\\
2008 Oct 22 & 21.09$\pm$0.04 & 17.77 & 0.71  & 0.50$\pm$0.05 & 0.31$\pm$0.05  & 0.36$\pm$0.05\\
2008 Nov 11 & 21.21$\pm$0.10 & 17.48 & 0.81 & -- & -- & -- \\
 &  &     & & &  &   \\
\hline
Mean Color &  &     & &0.63$\pm$0.03& 0.37$\pm$0.03  & 0.39$\pm$0.03  \\

Solar Color\tablenotemark{d} &  &    & & 0.63  & 0.36 & 0.33  \\
\enddata


\tablenotetext{a}{Apparent red magnitude within a 2.$\!\!^{\prime\prime}$2 radius aperture.}
\tablenotetext{b}{R-band magnitude corrected to unit heliocentric and geocentric distances and to zero phase angle, as described in the text.}
\tablenotetext{c}{Effective radius computed under the assumption of albedo 0.05.}
\tablenotetext{d}{Colors of the Sun, from Hardorp (1982), Hartmann et al.\ (1990) and Tueg and Schmidt-Kaler (1982) }

\end{deluxetable}

\clearpage 

\begin{deluxetable}{llcclllr}
\tabletypesize{\scriptsize}
\tablecaption{Summary of MBC Properties\tablenotemark{a} 
\label{summary}}
\tablewidth{0pt}
\tablehead{
\colhead{Object} & \colhead{$a$ (AU)\tablenotemark{b}} & \colhead{$e$\tablenotemark{c}}& \colhead{$i$ (deg)\tablenotemark{d}}   & \colhead{$T_J$\tablenotemark{e}} & \colhead{$q$ (AU)\tablenotemark{f}} & \colhead{$Q$ (AU)\tablenotemark{f}}  & $d_e$ (km)\tablenotemark{g}}
\startdata

133P/(7968) Elst-Pizarro& 3.156 & 0.165 & ~1.4 & 3.184 & 2.636 & 3.677 & 4.6  \cr

P/2005 U1 (Read) & 3.165 & 0.253 & ~1.3 & 3.153 & 2.365 & 3.965 & $<$0.6 \cr

176P/LINEAR (118401) & 3.196 & 0.192 & ~0.2 & 3.166 & 2.581 & 3.811 & 4.8 \cr

P/2008 R1 & 2.726 & 0.342 & 15.9 & 3.216 & 1.793 & 3.658 & $<$1.4 \cr

\enddata

\tablenotetext{a}{Comets with $T_J >$ 3.05 and $a < a_J$}
\tablenotetext{b}{Semimajor axis in AU}
\tablenotetext{c}{Orbital eccentricity}
\tablenotetext{d}{Orbital inclination (degrees)}
\tablenotetext{e}{Tisserand parameter with respect to Jupiter, by Equation (\ref{tisserand}) }
\tablenotetext{f}{Perihelion, $q$ and aphelion, $Q$, distances, in AU }
\tablenotetext{g}{Effective circular diameter, in km. }

\end{deluxetable}

\clearpage

\begin{figure}[]
\begin{center}
\includegraphics[width=0.8\textwidth]{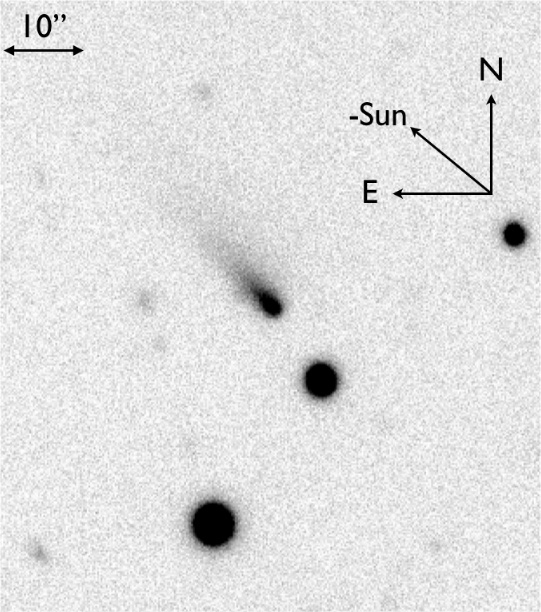}
\caption{R-band, 90 second image taken UT 2008 Sep 30 at the Keck telescope.  A tail of material extends to the north east from the nucleus of P/2008 R1.  North and south cardinal directions are marked, as is the projected Sun-comet radius vector (marked ``-Sun'').
\label{image_r1}}
\end{center} 
\end{figure}

\clearpage

\begin{figure}[]
\begin{center}
\includegraphics[width=0.8\textwidth]{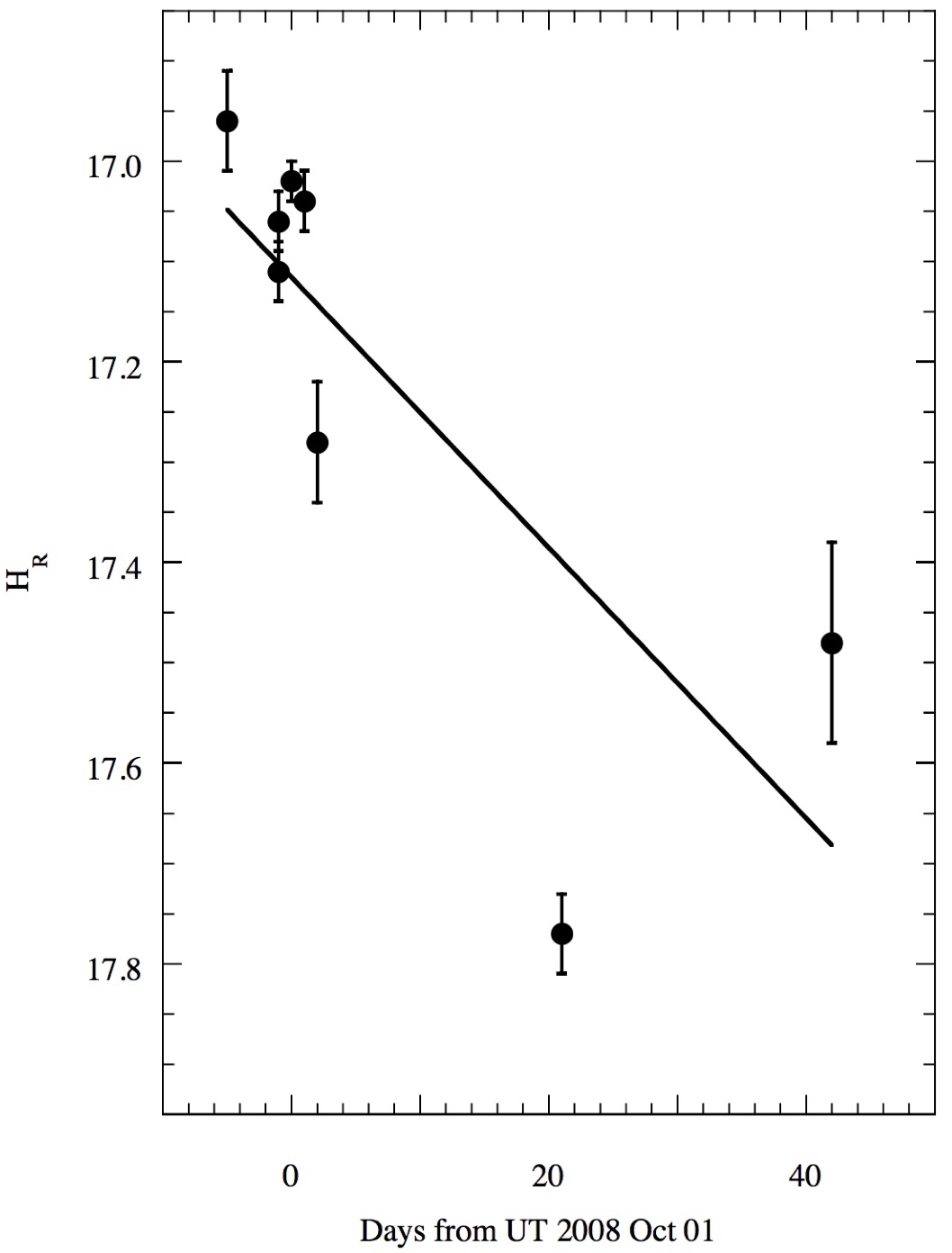}
\caption{Absolute red magnitude (i.e.\ corrected to unit heliocentric and geocentric distances and
to zero phase angle) of P/2008 R1 as a function of date (measured in days from UT 2008 Oct 01).
Error bars denote photometric errors, principally due to sky background uncertainty. \label{photometry}} 
\end{center} 
\end{figure}

\clearpage

\begin{figure}[]
\begin{center}
\includegraphics[width=0.9\textwidth]{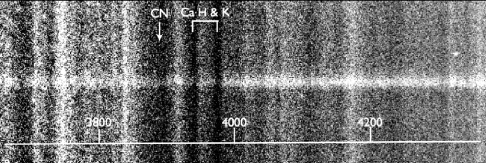}
\caption{The Keck spectrum of P/2008 R1 on UT 2008 Sep 30, before sky subtraction, with a wavelength scale in \AA~at the bottom.  The Solar Calcium H and K absorptions (back-scattered from Zodiacal dust) are shown and the position of the CN bandhead at 3883\AA~is indicated.  \label{2D}} 
\end{center} 
\end{figure}

\clearpage

\begin{figure}[]
\begin{center}
\includegraphics[width=0.95\textwidth]{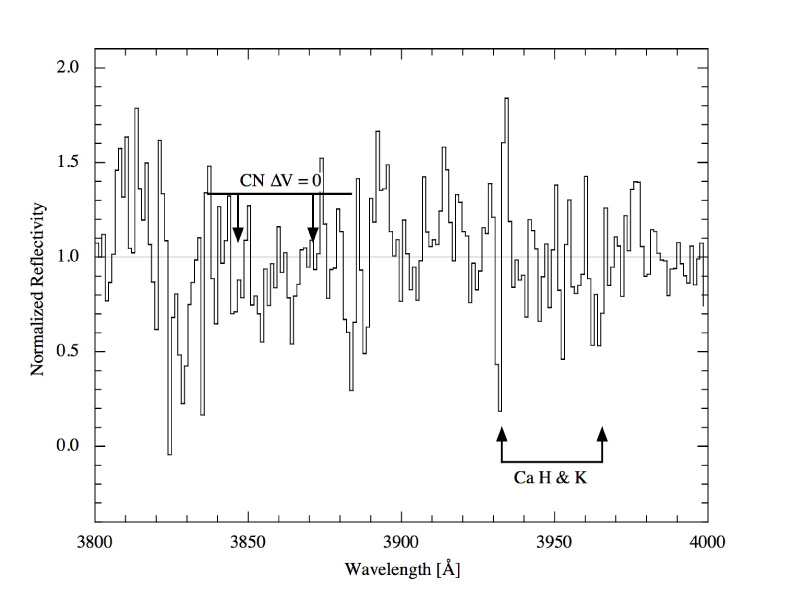}
\caption{Spectrum of P/2008 R1 taken UT 2008 Sep 30 at the Keck 10-m telescope.  This is the composite of four spectra (total
integration time 1710 s) that have been sky-subtracted, extracted from a region 1.$\!\!^{\prime\prime}$0 wide by 3.$\!\!^{\prime\prime}$8 long, and
divided by the spectrum of a nearby Solar analog to cancel photospheric absorption lines. The wavelengths of the Ca H and K lines are marked to show that they have been well cancelled by the Solar analog star and a horizontal line denotes the adopted upper limit to the CN $\Delta V$ = 0 band.\label{}} 
\end{center} 
\end{figure}

\clearpage

\begin{figure}[]
\begin{center}
\includegraphics[width=0.95\textwidth]{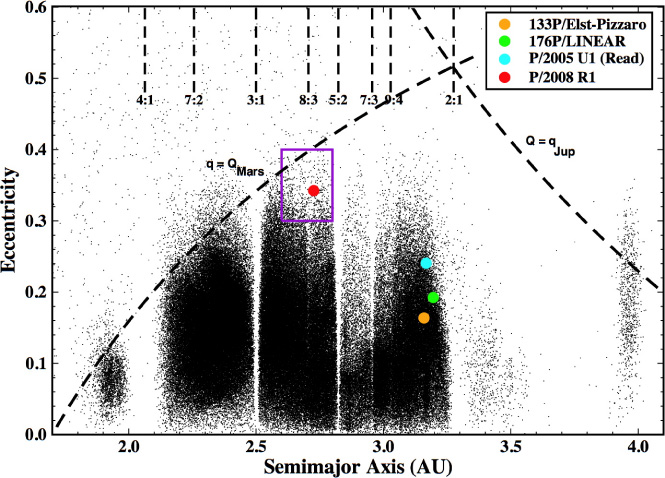}
\caption{The semimajor axis (AU) vs. orbital eccentricity plane showing  asteroids (black dots) and main-belt comets (color-coded as marked).   Orbital elements are from the JPL database.  The arcs show the loci of orbits having perihelia equal to the aphelion distance of Mars and aphelia equal to the perihelion distance of Jupiter.  The box around P/2008 R1 shows the region of $a$ vs.\ $e$ space in which dynamical stability is explored (see Figure 6).  \label{ae}} 
\end{center} 
\end{figure}

\clearpage

\begin{figure}[]
\begin{center}
\includegraphics[width=0.85\textwidth]{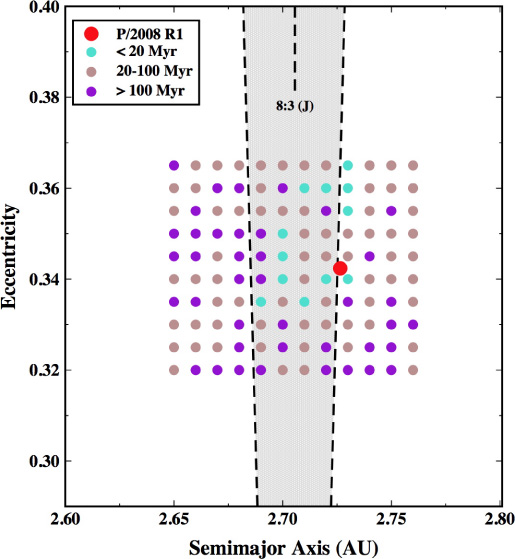}
\caption{Grid of clones in the semimajor axis vs.\ eccentricity plane for which the dynamical lifetime was calculated.  P/2008 R1 is shown as a red circle.  The grey band shows the location and width of the 8:3 mean-motion resonance with Jupiter. The lifetimes of the clones are shown with different colors.\label{clones}} 
\end{center} 
\end{figure}


\clearpage

\begin{thebibliography}{}

\bibitem[A'Hearn(1982)]{1982come.coll..433A} A'Hearn, M.~F.\ 1982, IAU 
Colloq.~61: Comet Discoveries, Statistics, and Observational Selection, 433 

\bibitem[A'Hearn et al.(1995)]{1995Icar..118..223A} A'Hearn, M.~F., Millis, 
R.~L., Schleicher, D.~G., Osip, D.~J., 
\& Birch, P.~V.\ 1995, Icarus, 118, 223

\bibitem[Arvesen et al.(1969)]{1969ApOpt...8.2215A} Arvesen, J.~C., 
Griffin, R.~N., \& Pearson, B.~D.\ 1969, \ao, 8, 2215 

\bibitem[Biver et 
al.(2002)]{2002EM&P...90....5B}Biver, N., et al.\ 2002, Earth Moon and Planets, 90, 5 

\bibitem[Bowell et al.(1989)]{1989aste.conf..524B} Bowell, E., Hapke, B., 
Domingue, D., Lumme, K., Peltoniemi, J., 
\& Harris, A.~W.\ 1989, Asteroids II, 524

\bibitem[Bottke et al.(2002)]{2002Icar..156..399B} Bottke, W.~F., 
Morbidelli, A., Jedicke, R., Petit, J.-M., Levison, H.~F., Michel, P., 
\& Metcalfe, T.~S.\ 2002, Icarus, 156, 399 

\bibitem[Bottke et al.(2006)]{2006AREPS..34..157B} Bottke, W.~F., Jr., 
Vokrouhlick{\'y}, D., Rubincam, D.~P., 
\& Nesvorn{\'y}, D.\ 2006, Annual Review of Earth and Planetary Sciences, 34, 157 

\bibitem[Bro{\v z} 
\& Vokrouhlick{\'y}(2008)]{2008MNRAS.390..715B} Bro{\v z}, M., \& Vokrouhlick{\'y}, D.\ 2008, \mnras, 390, 715 

\bibitem[Carusi et 
al.(1995)]{1995EM&P...68...71C} Carusi, A., Kres{\'a}k, {\v L}., \& Valsecchi, G.~B.\ 1995, Earth Moon and Planets, 68, 71 

\bibitem[Chambers(1999)]{1999MNRAS.304..793C} Chambers, J.~E.\ 1999, 
\mnras, 304, 793 


\bibitem[Drilling 
\& Landolt(2000)]{2000asqu.book..381D} Drilling, J.~S., \& Landolt, A.~U.\ 2000, Allen's Astrophysical Quantities, 381 




\bibitem[Fern{\'a}ndez et al.(2002)]{2002Icar..159..358F} Fern{\'a}ndez, 
J.~A., Gallardo, T., \& Brunini, A.\ 2002, Icarus, 159, 358

\bibitem[Fern{\'a}ndez et al.(2005)]{2005AJ....130..308F} Fern{\'a}ndez, 
Y.~R., Jewitt, D.~C., \& Sheppard, S.~S.\ 2005, \aj, 130, 308

\bibitem[Fornasier et al.(2007)]{2007Icar..190..622F} Fornasier, S., Dotto, 
E., Hainaut, O., Marzari, F., Boehnhardt, H., de Luise, F., 
\& Barucci, M.~A.\ 2007, Icarus, 190, 622 


\bibitem[Garradd et al.(2008)]{2008MPEC....S...46G} Garradd, G.~J., 
McNaught, R.~H., Meyer, M., Herald, D., 
\& Marsden, B.~G.\ 2008, Minor Planet Electronic Circulars, 46 

\bibitem[Garaud 
\& Lin(2007)]{2007ApJ...654..606G} Garaud, P., \& Lin, D.~N.~C.\ 2007, \apj, 654, 606 


\bibitem[Gomes et al.(2005)]{2005Natur.435..466G} Gomes, R., Levison, 
H.~F., Tsiganis, K., \& Morbidelli, A.\ 2005, \nat, 435, 466

\bibitem[Hag 2009]{hag09} Haghighipour, N.\ 2009, submitted 

\bibitem[Hardorp 1982]{har82} Hardorp, J.\ 1982, \aap, 105, 
120 

\bibitem[Hartmann et al.(1987)]{1987Icar...69...33H} Hartmann, W.~K., 
Tholen, D.~J., \& Cruikshank, D.~P.\ 1987, Icarus, 69, 33 

\bibitem[Hartmann et al.(1990)]{1990Icar...83....1H} Hartmann, W.~K., 
Tholen, D.~J., Meech, K.~J., \& Cruikshank, D.~P.\ 1990, Icarus, 83, 1 

\bibitem[Hsieh et al.(2004)]{2004AJ....127.2997H} Hsieh, H.~H., Jewitt, 
D.~C., \& Fern{\'a}ndez, Y.~R.\ 2004, \aj, 127, 2997 

\bibitem[Hsieh \& Jewitt(2006)]{2006Sci...312..561H} Hsieh, H.~H., \& Jewitt, D.\ 2006, Science, 312, 561 

\bibitem[]{hsi09} Hsieh, H., Jewitt, D., and Fernandez, Y.\ 2009.  Ap. J. Lett., submitted.

\bibitem[Jewitt(1991)]{1991ASSL..167...19J} Jewitt, D.\ 1991, Comets in the post-Halley Era, eds. R. Newburn, 
M. Neugebauer and J. Rahe. Kluwer Academic Publishers, Netherlands. pp. 19 - 65.

\bibitem[Jewitt(2004)]{2004come.book..659J} Jewitt, D.~C.\ 2004, COMETS II, edited by M. Festou, H. Weaver 
and U. Keller. Univ. Az. Press, Tucson. pp. 659 - 676.

\bibitem[Jewitt 
\& Meech(1986)]{1986ApJ...310..937J} Jewitt, D., \& Meech, K.~J.\ 1986, \apj, 310, 937 


\bibitem[Jones et al.(1990)]{1990Icar...88..172J} Jones, T.~D., Lebofsky, 
L.~A., Lewis, J.~S., \& Marley, M.~S.\ 1990, Icarus, 88, 172 
\bibitem[Kashikawa et al.(2002)]{2002PASJ...54..819K} Kashikawa, N., et 
al.\ 2002, \pasj, 54, 819 

\bibitem[Kosai(1992)]{1992CeMDA..54..237K} Kosai, H.\ 1992, Celestial 
Mechanics and Dynamical Astronomy, 54, 237

\bibitem[Kresak(1982)]{1982BAICz..33..104K} Kresak, L.\ 1982, Bulletin of 
the Astronomical Institutes of Czechoslovakia, 33, 104 

\bibitem[Landolt 1992]{1992AJ....104..340L} Landolt, A.~U.\ 1992, \aj, 
104, 340 

\bibitem[Levison et al.(2006)]{2006Icar..182..161L} Levison, H.~F., 
Terrell, D., Wiegert, P.~A., Dones, L., 
\& Duncan, M.\ 2006, Icarus, 182, 161

\bibitem[Meech  \& Jewitt(1987)]{1987A&A...187..585M} Meech, K.~J., \& Jewitt, D.~C.\ 1987, \aap, 187, 585

\bibitem[Millis et al.(1982)]{1982AJ.....87.1310M} Millis, R.~L., A'Hearn, 
M.~F., \& Thompson, D.~T.\ 1982, \aj, 87, 1310 

\bibitem[Nesvorn{\'y} et al.(2008)]{2008ApJ...679L.143N} Nesvorn{\'y}, D., 
Bottke, W.~F., Vokrouhlick{\'y}, D., Sykes, M., Lien, D.~J., 
\& Stansberry, J.\ 2008, \apjl, 679, L143

\bibitem[Oke et al.(1995)]{1995PASP..107..375O} Oke, J.~B., et al.\ 1995, 
\pasp, 107, 375

\bibitem[Rickman et al.(1991)]{1991AJ....102.1446R} Rickman, H., Kamel, L., 
Froeschle, C., \& Festou, M.~C.\ 1991, \aj, 102, 1446 

\bibitem[Roig et al.(2008)]{2008A&A...483..911R} Roig, F., Ribeiro, A.~O., \& Gil-Hutton, R.\ 2008, \aap, 483, 911 

\bibitem[Schorghofer(2008)]{2008ApJ...682..697S} Schorghofer, N.\ 2008, 
\apj, 682, 697

\bibitem[Tueg and Schmidt-Kaler 1982]{tue82} Tueg, H., \& 
Schmidt-Kaler, T.\ 1982, \aap, 105, 400 

\bibitem[Yoshikawa(1989)]{1989A&A...213..436Y} Yoshikawa, M.\ 1989, \aap, 213, 436 

\bibitem[Zappala et al.(1990)]{1990AJ....100.2030Z} Zappala, V., Cellino, 
A., Farinella, P., \& Knezevic, Z.\ 1990, \aj, 100, 2030 
\end{thebibliography}
\end{document}